\documentclass[aps,pra,showpacs,twocolumn,superscriptaddress]{revtex4}

\usepackage{graphicx,color}
\usepackage{amsmath,amsthm,amsfonts,amssymb,bm}
\usepackage{times}
\usepackage{epsf}
\usepackage[colorlinks={true}]{hyperref}

\usepackage[T1]{fontenc}
\usepackage[utf8]{inputenc}

\hypersetup{citecolor={blue}, filecolor={blue}, linkcolor={blue}, urlcolor={blue}}

\newcommand{\commentold}[1]{}
\DeclareMathSymbol{:}{\mathpunct}{operators}{"3A}

\begin{document}

\title{Non-Markovianity through flow of information between a system and an environment}

\author{S. Haseli}
\email{soroush.haseli@uok.ac.ir}
\affiliation{Department of Physics, University of Kurdistan, P.O.Box 66177-15175, Sanandaj, Iran}
\author{G. Karpat}
\affiliation{Faculdade de Ci\^encias, UNESP - Universidade Estadual Paulista, Bauru, SP, 17033-360, Brazil}
\author{S. Salimi}
\email{shsalimi@uok.ac.ir}
\affiliation{Department of Physics, University of Kurdistan, P.O.Box 66177-15175, Sanandaj, Iran}
\author{A.S. Khorashad}
\affiliation{Department of Physics, University of Kurdistan, P.O.Box 66177-15175, Sanandaj, Iran}
\author{F. F. Fanchini}
\email{fanchini@fc.unesp.br}
\affiliation{Faculdade de Ci\^encias, UNESP - Universidade Estadual Paulista, Bauru, SP, 17033-360, Brazil}
\author{B. \c{C}akmak}
\affiliation{Faculdade de Ci\^encias, UNESP - Universidade Estadual Paulista, Bauru, SP, 17033-360, Brazil}
\author{G. H. Aguilar}
\affiliation{Instituto de F\'{\i}sica, Universidade Federal do Rio de Janeiro, CP 68528, 21941-972, Rio de Janeiro, RJ, Brazil}
\author{S. P. Walborn}
\affiliation{Instituto de F\'{\i}sica, Universidade Federal do Rio de Janeiro, CP 68528, 21941-972, Rio de Janeiro, RJ, Brazil}
\author{P. H. Souto Ribeiro}
\email{phsr@if.ufrj.br}
\affiliation{Instituto de F\'{\i}sica, Universidade Federal do Rio de Janeiro, CP 68528, 21941-972, Rio de Janeiro, RJ, Brazil}

\date{\today}

\begin{abstract}
Exchange of information between a quantum system and its surrounding environment plays a fundamental role in the study of the dynamics of open quantum systems. Here we discuss the role of the information exchange in the non-Markovian behavior of dynamical quantum processes following the decoherence approach, where we consider a quantum system that is initially correlated with its measurement apparatus, which in turn interacts with the environment. We introduce a new way of looking at the information exchange between the system and environment using the quantum loss, which is shown to be closely related to the measure of non-Markovianity based on the quantum mutual information. We also extend the results of [Phys. Rev. Lett. 112, 210402 (2014)] by Fanchini et al. in several directions, providing a more detailed investigation of the use of the accessible information for quantifying the backflow of information from the environment to the system. Moreover, we reveal a clear conceptual relation between the entanglement and mutual information based measures of non-Markovianity in terms of the quantum loss and accessible information. We compare different ways of studying the information flow in two theoretical examples. We also present experimental results on the investigation of the quantum loss and accessible information for a two-level system undergoing a zero temperature amplitude damping process. We use an optical approach that allows full access to the state of the environment.
\end{abstract}


\maketitle

\section{Introduction}

The investigation of open quantum systems from various different perspectives has been subject of intense research in recent years motivated by fundamental questions, and also due to their crucial role in the realization of quantum information protocols in real world situations \cite{bbook,book2,book3}. One interesting approach to address open quantum systems is through the information flow among constituents of composite quantum systems, or in particular, to explore the exchange of information between the system of interest and its surrounding environment. From the point of view of memory effects, the dynamical quantum maps are usually divided in two groups, namely, Markovian and non-Markovian maps. Memoryless processes are often recognized as Markovian, where the information is expected to monotonically flow from the system to the environment. On the other hand, it is rather natural to assume that the backflow of information from the environment to the system is connected to the presence of memory effects, because in these cases the future states of the system may depend on its past states as a result of the inverse exchange of information.

Quantum Markovian maps are traditionally defined as the ones obtained from the solutions of Lindblad type master equations, which can be described by quantum dynamical semigroups \cite{bbook}. Thus, manifestation of memory effects in the form of recoherence and blackflow of information has been associated to the violation of the semigroup property. However, for such memory effects to emerge, failure to satisfy the semigroup property is not sufficient. The quantum map should also violate another property called divisibility \cite{piilo}. Recently, there has been an ever increasing interest in the non-Markovian nature of quantum processes \cite{resourece,various} and quantifying their degree of non-Markovianity using several distinct criteria \cite{rhp,hou,blp,luo,measures}. Whereas some authors have directly adopted the property of divisibility as the defining feature of quantum Markovian processes \cite{rhp,hou}, others have employed different means to identify memory effects \cite{blp,luo,measures}, which are not exactly equivalent but closely related to divisibility approach. In fact, unlike its classical counterpart, there is no universal definition of non-Markovianity in the quantum domain, and different measures do not coincide in general. Yet, it is reasonable to believe that conceptually different measures capture complementary aspects of the same phenomenon.

One of the most widely studied and significant quantifiers of the degree of non-Markovianity has been proposed by Breuer, Laine and Piilo (BLP) \cite{blp}. Rather than defining non-Markovianity based on the violation of divisibility, the BLP measure intends to determine the amount of non-Markovianity of a quantum process by checking the trace distance between two arbitrary states of the open system during the dynamics, which in fact quantifies the probability of successfully distinguishing these two states. Considering that the ability of distinguishing two objects is in a sense related to how much information we have about them, it is claimed that the monotonic reduction of distinguishability can be directly interpreted as a one-way flow of information from the system to the environment, which defines a Markovian quantum process. In contrast, if there is a temporary increase of trace distance throughout the time evolution of the system, then the quantum map is said to be non-Markovian due to the backflow of information from the environment to the system.

Another popular approach to quantify the degree of non-Markovianity is based on a well known property of local completely positive trace-preserving (CPTP) maps, that is, on their inability to increase entanglement between a system and an isolated ancilla \cite{mono}. Rivas, Huelga and Plenio (RHP) have introduced a witness for non-divisibility of quantum maps by making use of the monotonic behavior of entanglement measures under CPTP maps \cite{rhp}. Although this quantity does not provide a necessary and sufficient condition for divisibility, it can be adopted as a measure of non-Markovianity on its own since it has been shown that it encapsulates the information exchange between the system and environment through the concept of accessible information \cite{prl}. In this case, a non-Markovian process is characterized by a temporary increase of entanglement between the system and the isolated ancilla, which is an indicator of the backflow of information from the environment to the system. In the same spirit, Luo, Fu, and Song (LFS) have proposed a similar quantity that relies on the mutual information between a system and an arbitrary ancilla instead of entanglement \cite{luo}. Despite being easier to manage than entanglement-based measure mathematically, especially for high dimensional systems, this quantity does not yet have an interpretation directly related to the flow of information between the system and the environment.

In this work, our aim is twofold. First, using the language of the decoherence program, where a system $\mathcal{S}$ is coupled to a measurement apparatus $\mathcal{A}$, which in turn interacts with an environment $\mathcal{E}$ \cite{zurek}, we introduce a simple scheme to demonstrate how quantum loss \cite{qloss} can be utilized to describe the backflow of information from the environment $\mathcal{E}$ to the system $\mathcal{S}$. This approach is shown to be exactly equivalent to the LFS measure of non-Markovianity and thus gives it an interpretation in terms of information exchange between the system $\mathcal{S}$ and the environment $\mathcal{E}$. Furthermore, we reveal how the entanglement and the mutual information based measures of non-Markovianity are conceptually related to each other through the connection between quantum loss and accessible information. Second, we extend the results of Ref. \cite{prl} in several new directions. In particular, we investigate the role of both accessible information and quantum loss in quantifying non-Markovian behavior, via conceptually different means of information flow, in two paradigmatic models. We find out that, unlike the BLP measure, both LFS and RHP measures can capture the dynamical information in the non-unital aspect of the dynamics and thus can successfully identify non-Markovian behavior in corresponding models. Lastly, for a two-level system undergoing relaxation at zero temperature, we experimentally demonstrate the connection between quantum loss and quantum mutual information performing a quantum simulation with an all optical setup that allows full access to the environmental degrees of freedom \cite{exp}.

This paper is organized as follows. In Sec.II  we introduce the definitions of the considered non-Markovianity measures, and discuss how they are related to the flow of information between the system and the environment. We also present a clear conceptual connection between the quantum loss and the accessible information in quantifying information exchange. In Sec. III, using the RHP, LFS and BLP measures, we examine two examples of paradigmatic quantum channels theoretically, and present the experiment and its results. Section IV includes the discussion and summary of our findings.

\section{Measuring non-Markovianity via information flow}

Let us first define the type of quantum processes that we consider. We assume that a dynamical quantum map is described by a time-local master equation of the Lindblad form
\begin{equation}
\frac{\partial}{\partial t}\rho(t)=\mathcal{L}\rho(t), \label{master}
\end{equation}
with the Lindbladian super-operator $\mathcal{L}$ \cite{lindblad} given as
\begin{eqnarray}
\mathcal{L}\rho&=&-i[H,\rho] +\sum_{i}\gamma_{i}\left[A_i\rho A_i^\dagger-\frac{1}{2}\left\{A_i^\dagger A_i,\rho\right\}\right],\nonumber
\end{eqnarray}
where $H$ is the Hamiltonian of the system, $\gamma_i$'s are the decay rates, and $A_i$'s are the Lindblad operators describing the type of noise affecting the system. Provided $A_i$'s and $\gamma_i$'s are time independent, and also all $\gamma_i$'s are positive, Eq. (\ref{master}) leads to a dynamical semigroup of CPTP maps $\Lambda(t,0)=\exp[\mathcal{L}t]$ with $t>0$ satisfying the semigroup property
\begin{equation}
\Lambda(t_1+t_2,0)=\Lambda(t_1,0)\Lambda(t_2,0),
\end{equation}
for all $t_1,t_2\geq0$. Such a quantum dynamics defines a conventional Markovian process. However, it is possible that the Hamiltonian $H$, noise operators $A_i$, and decay rates $\gamma_i$ may have explicit time dependence. In this case, Eq. (\ref{master}) leads to what is known as a time-dependent Markovian process, if $\gamma_i(t)\geq0$ throughout the time evolution of the system. Dynamical maps might be written in terms of a time-ordered exponential as $\Lambda(t,0)=T \exp [\int_{0}^{t}\mathcal{L}(t')dt']$, which takes the state at time $0$ to the state at time $t$. Such Markovian maps have a fundamental property that they satisfy the condition of divisibility. In particular, a CPTP map $\Lambda(t_2,0)$ can be expressed as a composition of two other CPTP maps as
\begin{equation}
\Lambda(t_2,0)=\Lambda(t_2,t_1)\Lambda(t_1,0) \label{div}
\end{equation}
with $\Lambda(t_2,t_1)=T \exp [\int_{t_1}^{t_2}\mathcal{L}(t')dt']$, for all $t_1,t_2\geq0$. It is important to stress that time-dependent decay rates $\gamma_i(t)$ may become temporarily negative during the dynamics of the system. In such a situation, there exists an intermediate dynamical map $\Lambda(t_2,t_1)$ which is not CPTP, and thus violating the composition law for divisibility given by Eq. (\ref{div}) \cite{postvrates}. Let us remember that what we have introduced as a time-dependent Markovian process above is centered on the property of divisibility. However, the criteria for non-Markovian dynamics that we will discuss in the following sections are not exactly equivalent to non-divisibility of the dynamical maps, but rather rely on the idea of information backflow from the environment to the system from three conceptually different points of view.

\subsection{Trace Distance}

The BLP measure of non-Markovianity \cite{blp} is constructed upon the trace distance between two arbitrary states $\rho_1(t)$ and $\rho_2(t)$ of the reduced system of interest, which is given by
\begin{equation}
D(\rho_1(t),\rho_2(t))=1/2 \rm{Tr}|\rho_1(t)-\rho_2(t)|,
\end{equation}
where $|A|=\sqrt{A^\dagger A}$. It has been discussed that the trace distance has a physical interpretation in terms of the relative distinguishability of two quantum states. Suppose that Alice prepares a quantum system in either $\rho_1$ or $\rho_2$, with equal probabilities, and then sends it to Bob whose goal is to perform a single measurement on the system to reveal its state. In this scenario, it is possible to show that Bob can successfully identify the state of the system with an optimal probability of $1/2[1+D(\rho_1,\rho_2)]$. Hence, the trace distance can in fact be thought as a quantifier of the distinguishability of two states, variation of which during the evolution can be interpreted as an information exchange between the system and the environment. In particular, a monotonic loss of distinguishability between $\rho_1(t)$ and $\rho_2(t)$ during the dynamics, i.e. $dD(t)/dt<0$, indicates that information flows from the system to the environment at all times and thus the process is Markovian. On the other hand, $dD(t)/dt>0$ means that there exists a backlow of information from the environment to the system, giving rise to a non-Markovian process. Based on this criterion, the BLP measure is defined as
\begin{eqnarray}
\mathcal{N}_{BLP}(\Lambda)&=&\max_{\rho_1(0),\rho_2(0)}\int_{(dD(t)/dt)>0}\frac{dD(t)}{dt}dt
\end{eqnarray}
where the maximum is taken over all possible pairs of initial states $\rho_1(0)$ and $\rho_2(0)$. We should also note that the above equation can also be equivalently expressed as
\begin{equation}
\mathcal{N}_{BLP}(\Lambda)=\max_{\rho_1(0),\rho_2(0)}\sum_i [D(b_i)-D(a_i)],\label{Sblpm}
\end{equation}
where time intervals $(a_i,b_i)$ correspond to the regions where $dD(t)/dt>0$, and maximization is done over all pairs of initial states. Even though this optimization is a considerably hard task, it is possible to simplify the procedure in several ways by reducing the number of possible optimizing pairs.

Furthermore, considering the fact that CPTP maps are contractions for the trace distance, one can show that the distinguishability between $\rho_1(t)$ and $\rho_2(t)$ is guaranteed to monotonically decrease for all divisible processes. Thus, according to the BLP measure, all divisible dynamical maps define Markovian processes. Nonetheless, the inverse statement is not necessarily true, that is, there exist non-divisible maps for which the trace distance does not show any temporary revival at all. In fact, the trace distance is actually a witness for non-divisibility. In addition, it has been recently shown that the trace distance is not able to capture the dynamical information in the non-unital aspect of quantum dynamics. Consequently, it fails to identify the non-Markovianity originated from the non-unital part of the transformation \cite{nonunital}.

\subsection{Quantum loss}

\begin{figure}[t]
\includegraphics[width=0.47\textwidth]{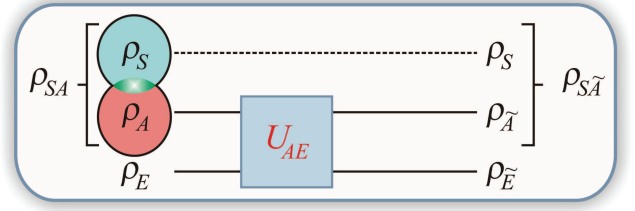}
\caption{We consider an initially pure environment $\mathcal{E}$, and an entangled pure state $\mathcal{SA}$. As the system $\mathcal{S}$ evolves free of any direct interaction, the apparatus is interacting with the environment $\mathcal{E}$.}
\label{fig1}
\end{figure}

In this section, we introduce a new method of quantifying non-Markovianity through the flow of information between the system and the environment, using a conceptually different approach than the BLP measure. Our discussion relies on the decoherence program, where a quantum system $\mathcal{S}$ is coupled to a measurement apparatus $\mathcal{A}$, which in turn directly interacts with an environment $\mathcal{E}$. Let us first consider a quantum system $\mathcal{S}$ that is initially correlated with the apparatus $\mathcal{A}$. We assume that the bipartite system $\mathcal{SA}$ starts as a pure state, and the environment only affects the state of the apparatus $\mathcal{A}$. As a result of the interaction, there emerges an amount of correlation among the individual parts of the closed tripartite system $\mathcal{SAE}$, and thus the environment $\mathcal{E}$ acquires information about the system $\mathcal{S}$ by means of the interaction with the apparatus $\mathcal{A}$. This setting is graphically sketched in Fig. \ref{fig1}, where the system $\mathcal{S}$ evolves trivially while the apparatus $\mathcal{A}$ is in a direct unitary interaction with the environment $\mathcal{E}$. The final state of the composite tripartite system $\mathcal{SAE}$ is given by
\begin{equation}
\rho_{\mathcal{S\tilde{A}\tilde{E}}}=(I^{\mathcal{S}} \otimes U^{\mathcal{AE}})\rho_{\mathcal{SAE}}(I^{\mathcal{S}} \otimes U^{\mathcal{AE}})^{\dag},
\end{equation}
where tilde denotes the state of the subsystems after the time evolution. The resulting state of each part of the composite system can be obtained by tracing over the remaining parts. Particularly, if we discard the environment $\mathcal{E}$, we obtain the bipartite state of the system $\mathcal{S}$ and the apparatus $\mathcal{A}$ as
\begin{equation}
\rho_{\mathcal{S\tilde{A}}}=\rm{Tr}_{\mathcal{E}}(\rho_{\mathcal{S\tilde{A}\tilde{E}}}),
\end{equation}
which corresponds to applying a general CPTP map to the apparatus $\mathcal{A}$ while leaving the state of the system $\mathcal{S}$ untouched.

Let us now introduce some preliminary concepts that will be relevant to our treatment of the information exchange between the system and the environment. For a bipartite system $XY$, while the conditional quantum entropy is defined as
\begin{equation}
S(X|Y)=S(XY)-S(Y),
\end{equation}
the quantum mutual information is given by
\begin{align}
S(X:Y) &= S(X)-S(X|Y) \\
       &= S(X)+S(Y)-S(XY), \nonumber
\end{align}
where $S(X(Y))\equiv S\left(\rho_{X(Y)}\right)=-\rm{Tr}\left(\rho_{X(Y)}\log_{2}\rho_{X(Y)}\right)$ denotes the von-Neumann entropy of the considered systems, characterizing the uncertainty about them. Provided that we have a quantum system $XY$ in a pure state, we obtain $S(XY)=0$ and, as a result, $S(X:Y)=2S(X)=2S(Y)$. Moving to tripartite system $XYZ$, we can define the conditional quantum entropy of $X$ and $Y$ conditionally on Z as
\begin{align}
S(X:Y|Z) &= S(X|Z)-S(X|YZ) \\
         &= S(X|Z)+S(Y|Z)-S(XY|Z)  \nonumber \\
         &= S(XZ)+S(YZ)-S(Z)-S(XYZ), \nonumber
\end{align}
whereas the quantum ternary mutual information reads
\begin{equation}
S(X:Y:Z)=S(X:Y)-S(X:Y|Z).
\end{equation}
It is important to note that $S(X:Y|Z)\geq0$, and the quantum ternary mutual information $S(X:Y:Z)$ vanishes for a pure tripartite state, i.e., $S(X:Y)=S(X:Y|Z)$.

\begin{figure}[t]
\includegraphics[width=0.48\textwidth]{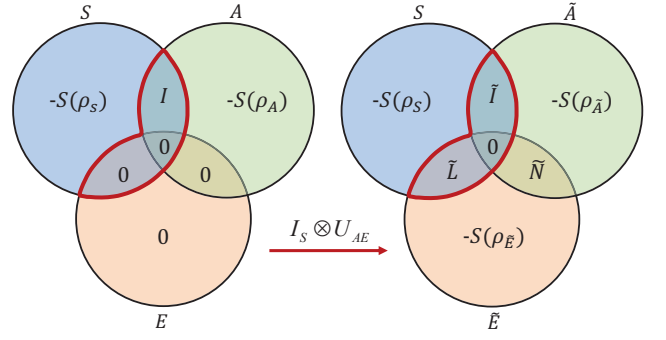}
\caption{The entropy diagram of the tripartite system composed of the system $\mathcal{S}$, the apparatus $\mathcal{A}$, and the environment $\mathcal{E}$, before and after the interaction. The amount of information will stay the same inside the area enclosed by thick red curves, i.e., $I=\tilde{I}+\tilde{L}=2S(\rho_{\mathcal{S}})$, where $\tilde{I}$ is the mutual information and $\tilde{L}$ is the quantum loss.}
\label{fig2}
\end{figure}

In the following, we adopt the terminology introduced in Ref. \cite{qloss}. Making use of the analogy with the classical information theory, we can make an entropy diagram for the composite system of $\mathcal{SAE}$ to show how each part of the tripartite system share and exchange information among its subsystems. For this purpose, we define three quantities that will be very useful to describe the information dynamics of the tripartite system, namely, the quantum mutual information $\tilde{I}$, the quantum loss $\tilde{L}$, and the quantum noise $\tilde{N}$:
\begin{align}
\tilde{I} &= S(\mathcal{S:\tilde{A}}), \\
\tilde{L} &= S(\mathcal{S:\tilde{E}|\tilde{A}})=S(\mathcal{S:\tilde{E}}), \\
\tilde{N} &= S(\mathcal{\tilde{A}:\tilde{E}|S})=S(\mathcal{\tilde{A}:\tilde{E}}).
\end{align}
In Fig. \ref{fig2}, we display the entropy diagram for $\mathcal{SAE}$ using the above quantities. The quantum mutual information $\tilde{I}$ quantifies the amount of residual mutual entropy between the system $\mathcal{S}$ and the apparatus $\mathcal{A}$ after the decoherence occurs. On the other hand, the quantum loss $\tilde{L}$ represents the amount of information that is getting lost in the environment $\mathcal{E}$. Actually, among these three quantities, only $\tilde{I}$ and $\tilde{L}$ are relevant to us, since information exchange between the system $\mathcal{S}$ and the environment $\mathcal{E}$ is characterized by the balance between them. It is very important to emphasize that the equality
\begin{equation} \label{IL}
\tilde{I}+\tilde{L}=2S(\rho_{\mathcal{S}})=2S(\rho_{\mathcal{A}})
\end{equation}
holds at all times during the dynamics. That is, twice the initial entropy of the system $\mathcal{S}$ will be redistributed to the apparatus $\mathcal{A}$ and the environment $\mathcal{E}$ as decoherence takes place. In other words, the total amount of information inside the closed thick red line in Fig. \ref{fig2} will remain invariant. Indeed, as the composite tripartite system $\mathcal{SAE}$ evolves in time, the environment $\mathcal{E}$ will learn about the system $\mathcal{S}$, and the quantum mutual information $I$ will start to decrease as a result of its monotonicity under local CPTP maps. This will be directly reflected as an increase in the quantum loss $\tilde{L}$, which is naturally zero initially, as can be observed from Eq. (\ref{IL}). However, it is also possible that $\tilde{I}$ might temporarily revive during dynamics, which will give rise to a temporary decrease in $\tilde{L}$.

Regarding non-Markovianity as a phenomenon that is intrinsically related to the backflow of information from the environment $\mathcal{E}$ to the system $\mathcal{S}$, it is reasonable to expect the quantum loss $\tilde{L}$ to monotonically increase for Markovian dynamics, since it is an entropic measure of information that the environment $\mathcal{E}$ acquires about the system $\mathcal{S}$. Therefore, one can define non-Markovian processes as the ones for which there is a temporary loss of $\tilde{L}$ as the system evolves in time, i.e. $d\tilde{L}/dt<0$, since this is an indication that the information flows back to the system $\mathcal{S}$ from the environment $\mathcal{E}$.

We should clarify that when we say information flow from the system $\mathcal{S}$ to the environment $\mathcal{E}$ or vice versa, we do not actually mean that total information content of the system changes, since it is constant at all times due to the fact that the system $\mathcal{S}$ does not directly interact with the environment $\mathcal{E}$. Rather, we mean that information is being redistributed in the tripartite composite system $\mathcal{SAE}$ in such a way that the amount of information that the system $\mathcal{S}$ shares with the environment $\mathcal{E}$ increases or decreases, as depicted in Fig. \ref{fig2}.

At first sight, one might think that evaluation of the quantum loss $\tilde{L}$ requires the knowledge of the state of the environment $\mathcal{E}$, which typically consists of infinite number of degrees of freedom, and is virtually impossible to access in real world situations. However, we actually do not need to directly access the environment to be able to calculate $\tilde{L}$. It can be explicitly written without dependence on the environment $\mathcal{E}$,
\begin{align}
\tilde{L} = S(\rho_{\mathcal{A}})-S(\rho_{\mathcal{\tilde{A}}})+S(\rho_{\mathcal{S\tilde{A}}}).
\end{align}

An interesting point is that the quantum loss $\tilde{L}$ can also be rewritten as a difference of the initial and final mutual information shared by the system $\mathcal{S}$ and the apparatus $\mathcal{A}$, that is,
\begin{equation} \label{LII}
\tilde{L}= I- \tilde{I},
\end{equation}
as can be easily seen from Eq. ({\ref{IL}}), since the initial mutual information $I$ is twice the initial entropy of the system $\mathcal{S}$. Taking the time derivative of this simple equation, we find that
\begin{equation} \label{dLI}
\frac{d}{dt}\tilde{L}= - \frac{d}{dt}\tilde{I}.
\end{equation}
Recalling that the LFS measure \cite{luo} of non-Markovianity is based on the rate of change of the quantum mutual information shared by the system $\mathcal{S}$ and the apparatus $\mathcal{A}$, we immediately realize that in fact the quantum loss approach to non-Markovianity is exactly equivalent to the formulation of the LFS measure. In particular, the LFS measure captures the non-Markovian behavior through a temporary increase of the mutual information of the bipartite system $\mathcal{SA}$.  Mathematically, the LFS measure can be written as
\begin{equation} \label{LFSmeas}
\mathcal{N}_{LFS}(\Lambda)=\max_{\rho_{\mathcal{SA}}}\int_{(d/dt)\tilde{I}>0}\frac{d}{dt}\tilde{I}dt,
\end{equation}
where the maximization is evaluated over all possible pure initial states of the bipartite system $\mathcal{SA}$. Thus, the quantum loss gives an interpretation to the LFS measure in terms of information exchange between the system $\mathcal{S}$ and the environment $\mathcal{E}$, since any temporary loss of $\tilde{L}$ will be observed as a temporary revival of $\tilde{I}$ by the same amount. Note that it directly follows from the composition law of divisibility given in Eq. (\ref{div}) that the LFS measure vanishes for all divisible quantum processes due to the monotonicity of the mutual information under CPTP maps. We should still keep in mind that the inverse statement is not always true. Some non-divisible maps do not increase the mutual information, or equivalently, decrease the quantum loss at all.

\begin{figure}[t]
\includegraphics[width=0.3\textwidth]{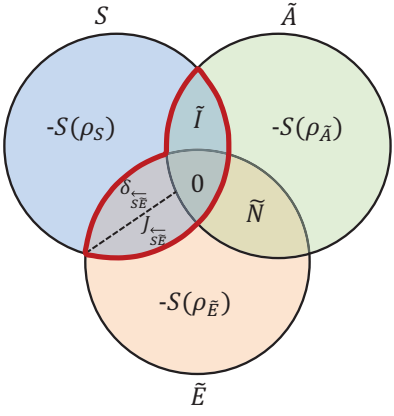}
\caption{The accessible information $J_{\mathcal{S\tilde{E}}}^\leftarrow$ and the inaccessible information $\delta_{\mathcal{S\tilde{E}}}^\leftarrow$ in the entropy diagram of the tripartite system $\mathcal{SAE}$ after the interaction of the apparatus $\mathcal{A}$ with the environment $\mathcal{E}$.}
\label{fig3}
\end{figure}

\subsection{Accessible Information}

Next, we introduce another quantity known as the accessible information \cite{hend}, which quantifies the maximum amount of classical information that can be extracted about the system $\mathcal{S}$ by locally observing the environment $\mathcal{E}$,
\begin{equation}
J_{\mathcal{SE}}^\leftarrow= \max_{\{\Gamma_{i}^{\mathcal{E}}\}} \left[S(\rho_{\mathcal{S}}) - \sum_i p_i S(\rho_{\mathcal{S}}^i|\Gamma_i^{\mathcal{E}})\right],
\label{ai}
\end{equation}
where $\{\Gamma_i^{\mathcal{E}}\}$ defines a complete positive operator valued measure (POVM) acting on the state of the environment $\mathcal{E}$, and $\rho_{\mathcal{S}}^{i}=\textmd{Tr}_{\mathcal{E}}((I^{\mathcal{S}}\otimes\Gamma_{i}^{\mathcal{E}})\rho_{\mathcal{SE}})/p_{i}$ is the remaining state of subsystem $\mathcal{S}$ after obtaining the outcome $i$ with the probability $p_{i}=\textmd{Tr}((I^{\mathcal{S}}\otimes\Gamma_{i}^{\mathcal{E}})\rho_{\mathcal{SE}})$. Considering the fact that the quantum loss $\tilde{L}$ is nothing but the quantum mutual information between the system $\mathcal{S}$ and the final state of the environment after decoherence, $\mathcal{\tilde{E}}$, it is possible to express it as
\begin{equation}
\tilde{L}=J_{\mathcal{S\tilde{E}}}^\leftarrow + \delta_{\mathcal{S\tilde{E}}}^\leftarrow,
\end{equation}
where $\delta_{\mathcal{S\tilde{E}}}^\leftarrow$ (known as the quantum discord in literature as a genuine measure of non-classicality \cite{ollivier}) quantifies the part of the quantum mutual information $\tilde{L}$ that the environment $\mathcal{\tilde{E}}$ cannot access about the system $\mathcal{S}$ locally during the decoherence process. In other words, despite the system $\mathcal{S}$ and the environment $\mathcal{\tilde{E}}$ have the information $\tilde{L}$ in common, we can only access a fraction of it, namely $J_{\mathcal{S\tilde{E}}}^\leftarrow$, by just observing the state of the environment $\mathcal{E}$. In Fig. \ref{fig3}, we display the accessible and and inaccessible information in the entropy diagram of $\mathcal{SAE}$ after the interaction starts to take place.

Returning to the discussion of non-Markovianity, one might argue that it is also quite reasonable to define non-Markovianity in terms of information flow using the accessible information instead of the quantum loss. The reason is that the accessible information measures the fraction of information that the environment $\mathcal{E}$ can actually access about the system $\mathcal{S}$, rather than the total amount of information they share as quantified by the quantum loss. Similarly to the case of $\tilde{I}$, we do not need any information about the state of the environment $\mathcal{S}$ to be able to evaluate the accessible information. Remembering that the environment $\mathcal{E}$ is initially in a pure state, that is, we consider a zero temperature reservoir, and also that the tripartite state $\mathcal{SAE}$ stays pure at all times, the Koashi-Winter relation implies that \cite{kw}
\begin{equation}
E_{\mathcal{SA}}=S(\rho_\mathcal{S})-J_{\mathcal{SE}}^\leftarrow.
\label{kw}
\end{equation}
where $E_{\mathcal{SA}}$ denotes the entanglement of formation shared by the system $\mathcal{S}$ and the apparatus $\mathcal{A}$, which is a resource-based measure quantifying the cost of generating a given state by means of maximally entangled resources \cite{eof}. It is given by
\begin{align}
             E(\rho) &= h\left( \frac{1+\sqrt{1-C^{2}(\rho)}}{2}\right); \\
             h(x) &= -x\log{x} -(1-x) \log{(1-x)},
\end{align}
where $C(\rho)=\max \left\{ 0,\sqrt{\lambda_{1}}-\sqrt{\lambda_{2}}-\sqrt{\lambda_{3}}-\sqrt{\lambda_{4}},\right\}$ with $\{\lambda_{i}\}$ being the eigenvalues of the product matrix $\rho \tilde{\rho}$ in decreasing order. Here, $\tilde{\rho}=(\sigma^{y}\otimes\sigma^{y})\rho^{*}(\sigma^{y}\otimes\sigma^{y})$, $\sigma^{y}$ is the Pauli spin operator in y-direction, and $\rho^{*}$ is obtained from $\rho$ via complex conjugation. Since the system $\mathcal{S}$ does not interact directly with the environment $\mathcal{E}$, we know that its state is invariant in time throughout the dynamics, then the time derivative of the Koashi-Winter relation given in Eq. (\ref{kw}) leads to a simple relation between the rate of changes of the entanglement of formation and the accessible information \cite{prl},
\begin{equation}
\frac{d}{dt}E_{\mathcal{S\tilde{A}}}=-\frac{d}{dt}J_{\mathcal{S\tilde{E}}}^\leftarrow.
\label{ej}
\end{equation}
This relation immediately implies that any temporary decrease in $J_{\mathcal{S\tilde{E}}}^\leftarrow$ will be reflected as a temporary increase of $E_{\mathcal{S\tilde{A}}}$. Thus, the non-Markovianity measure based on the rate of change of the accessible information $J_{\mathcal{SE}}^\leftarrow$ can also be expressed in terms of the rate of change of the entanglement of formation between the system $\mathcal{S}$ and the apparatus $\mathcal{A}$ \cite{prl}. At this point, we recall that the basis of the entanglement-based RHP measure of non-Markovianity \cite{rhp} is the monotonic behavior of entanglement measures under local CPTP maps. In particular, according to the RHP criterion, any temporary revival of entanglement is an indication of the non-Markovian nature of a quantum process. The RHP measure depends on the rate of change of the entanglement shared by the system $\mathcal{S}$ and the apparatus $\mathcal{A}$, and thus can be written as
\begin{equation}
\mathcal{N}_{RHP}(\Lambda)=\max_{\rho_{\mathcal{SA}}}\int_{(d/dt)E_{\mathcal{S\tilde{A}}}>0}\frac{d}{dt}
E_{\mathcal{S\tilde{A}}} dt, \label{measure}
\end{equation}
where the maximization is evaluated over all possible pure initial states of the bipartite system $\mathcal{SA}$. With the help of Eq. (\ref{ej}), it is now straightforward to observe that the entanglement-based RHP measure of non-Markovianity is indeed exactly equivalent to the accessible information approach. In other words, when entanglement of formation is chosen as a measure of entanglement, the RHP measure quantifies the total amount of decrease in the information that the environment $\mathcal{E}$ can access about the system $\mathcal{S}$. The composition law of divisibility given in Eq. (\ref{div}) implies that the RHP measure vanishes for all divisible quantum processes, just as in the case of the LFS measure, and again the inverse statement might not be always true since it is possible for some non-divisible maps not to increase entanglement.

It becomes clear with our interpretation that the LFS and RHP measures of non-Markovianity, despite being based on different physical quantities, are closely related to each other conceptually when the flow of information between the system $\mathcal{S}$ and the environment $\mathcal{E}$ is considered. From this angle, the only difference between them is the local accessibility of the information, that environment $\mathcal{E}$ and the system $\mathcal{S}$ have in common, by observing the environment $\mathcal{E}$. Especially, investigating the problem of information exchange from the point of view of the decoherence program, we demonstrate that both the mutual information and entanglement are relevant quantities for quantifying non-Markovianity as a backflow of information from the environment $\mathcal{E}$ to the system $\mathcal{S}$.

\begin{figure}[t]
\includegraphics[width=0.43\textwidth]{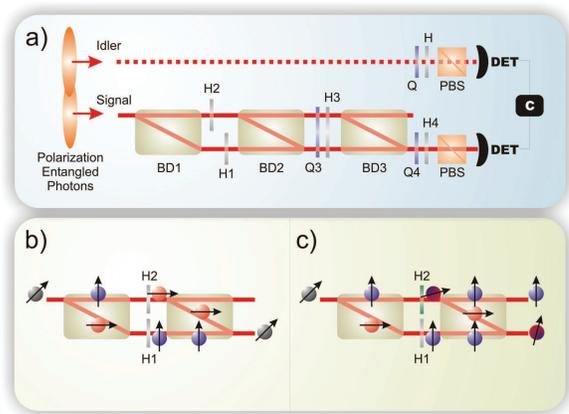}
\caption{(a) Sketch of the experimental set-up. The H's are half wave plates, the Q's are
quarter wave plates, the BD's are beam displacers, PBS is polarizing beam splitter, and DET is single photon detector. The light beams propagate from left to right. (b) Implementation of the amplitude damping channel for the photon polarization, condition $p=0$. (c) Implementation of the amplitude damping channel for the photon polarization, condition $p \neq 0$.}
\label{fig4}
\end{figure}

Getting back to the optimization of the LFS and the RHP measures, we should emphasize that it is in fact not necessary to perform the optimization over all variables appearing in the pure bipartite density matrix of $\mathcal{SA}$. We can actually simplify the optimization procedure for both LFS and RHP measures without loss of any generality as follows. For instance, in case of a pure two-qubit system, one can consider a general mixed single qubit density matrix for the apparatus $\mathcal{A}$, having only three real parameters, and then purify it to obtain the two-qubit pure state of the bipartite system $\mathcal{SA}$. It is known that all purifications of the apparatus $\mathcal{A}$ can be obtained by applying unitary operations locally on the system $\mathcal{S}$. Also note that the entanglement and quantum mutual information of $\mathcal{SA}$ remain invariant under these operations. Additionally, taking into account that the system $\mathcal{S}$ does not directly interact with the environment $\mathcal{E}$, the simplification is justified and three real variables are sufficient to perform the optimization.

\begin{figure*}[t]
\includegraphics[width=0.75\textwidth]{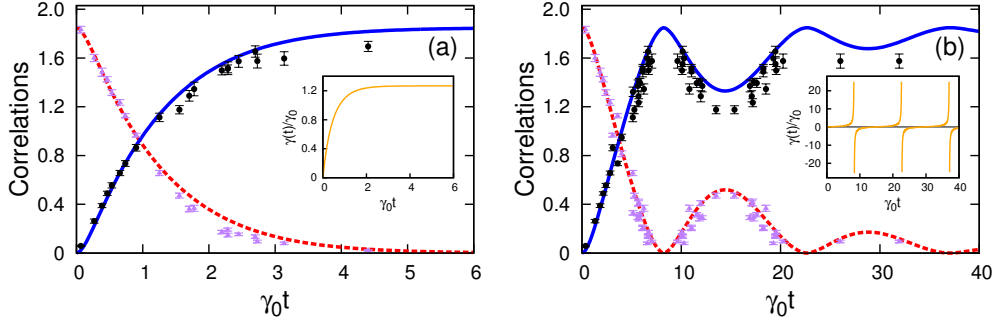}
\caption{Theoretical plot of the quantum loss $\tilde{L}$ (blue line) and the quantum mutual information $\tilde{I}$ (dashed red line). The insets display the decay rate $\gamma(t)/\gamma_0$ (orange line) as a function of scaled time $\gamma_0t$, that is being experimentally controlled by $p(t) \rightarrow \theta_p$. Experimental points for $\tilde{L}$ and $\tilde{I}$ are shown by black dots and purple triangles, respectively. As (a) demonstrates the monotonicity of information flow in the Markovian regime with $\lambda/\gamma_0=3$, (b) displays its non-monotonous behavior in the non-Markovian regime with $\lambda/\gamma_0=0.1.$}
\label{fig5}
\end{figure*}

Besides, note that we have assumed the environment $\mathcal{E}$ to be initially in a pure state. This assumption does not hold in general, in particular, when we consider a finite temperature environment. In this case, the initial state of the environment $\mathcal{E}$ is mixed, and the Koashi-Winter relation given in Eq. (\ref{kw}) becomes an inequality. However, we can purify the state of the environment $\mathcal{E}$ by extending the Hilbert space with a complementary subsystem $\mathcal{E'}$, without loss of generality. Consequently, we can again use the Koashi-Winter relation, which gives $E_{\mathcal{SA}}=S_\mathcal{S}-J^\leftarrow_{\mathcal{S}\{\mathcal{EE'}\}}$. We now see that the entanglement between the system $\mathcal{S}$ and the apparatus $\mathcal{A}$ is still connected to the information that the bipartite system $\mathcal{EE'}$ can access about the system $\mathcal{S}$. It is rather straightforward to see that a similar treatment can be done for the case of the quantum loss and the LFS measure via the extension of the environment $\mathcal{E}$ with an extra purifying system $\mathcal{E'}$.

\section{Examples}

In this section, we discuss the similarities and differences between the distinct ways of quantifying non-Markovianity based on information exchange between the system $\mathcal{S}$ and the environment $\mathcal{E}$, considering two relaxation models for open quantum systems. First, we examine the zero temperature relaxation channel, for which we present an all optical experimental simulation that realizes the required scenario to investigate the information flow in terms of the quantum loss and the accessible information. Second, we theoretically examine the generalized amplitude damping channel. We show that in this context there exist differences between the accessible information and quantum loss approaches.

\subsection{Amplitude Damping}

Here we treat the apparatus $\mathcal{A}$ as a two-level quantum system interacting with a zero temperature relaxation environment described by a collection of bosonic oscillators. The corresponding interaction Hamiltonian is given by
\begin{equation}
H=\omega_0\sigma_{+}\sigma_{-}+\sum_{k}\omega_k a_k^\dagger a_k + (\sigma_{+}B + \sigma_{-}B^\dagger)\label{jc},
\end{equation}
where $\sigma_{\pm}$ denote the raising and lowering operators of the apparatus $\mathcal{A}$ having the transition frequency $\omega_0$, and $B=\sum_k g_k a_k$. The annihilation and creation operators of the environment $\mathcal{E}$ are represented by $a_k$ and $a_k^\dagger$, respectively, with the frequencies $\omega_k$. We assume that the environment $\mathcal{E}$ has an effective spectral density of the form $J(\omega)= \gamma_0 \lambda^2 / 2\pi[(\omega_0 - \omega)^2 + \lambda^2]$, where the spectral width of the coupling $\lambda$ is related to the correlation time of the environment $\tau_B$ via $\tau_B\approx 1/\lambda$. The parameter $\gamma_0$ is connected to the time scale $\tau_R$, over which the state of the system changes, by $\tau_R\approx 1/\gamma_0$. Dynamics of the apparatus $\mathcal{A}$, with this spectral density, can be described by a master equation having the form of Eq. (\ref{master}),
\begin{equation}
\frac{\partial}{\partial t}\rho(t)=\gamma(t)\left(\sigma_{-}\rho(t)\sigma_{+}-\frac{1}{2}\{\sigma_{+}\sigma_{-},
\rho(t)\}\right), \label{masterjc}
\end{equation}
where the time-dependent decay rate is given by
\begin{equation}
\gamma(t)=\frac{2\gamma_0\lambda\sinh{(dt/2)}}{d\cosh{(dt/2)}+\lambda\sinh{(dt/2)}},
\end{equation}
with $d=\sqrt{\lambda^2-2\gamma_0\lambda}$. Then, we can express the dynamics of the apparatus $\mathcal{A}$ in the Kraus operator representation as
\begin{equation}
\rho(t)= \Lambda(\rho(0))=\sum_{i=1}^{2} M_i(t) \rho(0) M_i^\dagger(t),
\end{equation}
where the corresponding Kraus operators $M_i(t)$ are
\begin{align}
 M_1(t) &= \begin{pmatrix} 1 & 0\\ 0 & \sqrt{1-p(t)} \end{pmatrix}, &
 M_2(t) &= \begin{pmatrix} 0 & \sqrt{p(t)}\\ 0 & 0 \end{pmatrix},\label{kraus}
\end{align}
satisfying the condition $\sum_{i=1}^{2} M_i^\dagger(t) M_i(t) = I$ for all values of $t$, and the parameter
$p(t)$ is given by
\begin{equation}
p(t)=1-e^{-\lambda t} \left[ \cosh{\left(\frac{dt}{2}\right)+\frac{\lambda}{d}\sinh{\left(\frac{dt}{2}\right)}}\right]^2\label{pt}.
\end{equation}

The scenario of a system $\mathcal{S}$ entangled with a measurement apparatus $\mathcal{A}$,
which interacts with the environment $\mathcal{E}$ in the form of an amplitude damping channel, can be
realized with polarization entangled photon pairs and interferometers \cite{exp}. A sketch of the experimental set-up is presented in Fig. \ref{fig4}a. We employ a widely used source of polarization entangled photons \cite{exp2}. We send the signal photon through the interferometers and the idler photon goes straight to polarization analysis and detection. The polarization state of the idler photon represents the system $\mathcal{S}$. The polarization state of the signal photon represents the apparatus $\mathcal{A}$. The signal photon enters the first interferometer that implements the amplitude damping channel. In this way the polarization state of the signal photon evolves in perfect analogy with the
spontaneous emission of a two-level atom. In the case of the atom, as time passes, the amplitude probability associated to the excited state decreases exponentially. In the case of the photon polarization, there is a parameter $p$ ranging from $0$ to $1$, which is equivalent to time, so that $p = 1$ is equivalent to $t \rightarrow \infty$. Therefore, the interferometer produces a
controlled decrease of the polarization component representing the excited state.
In Fig. \ref{fig4}b, we show what happens when the control parameter is $p=0$ and the input state is
$|\psi_{in}\rangle = (1/\sqrt{2})(|H\rangle+|V\rangle)$. The incoming
photon splits in two polarization components. The vertical one is represented by the  blue ball
and is transmitted. The horizontal one is represented by the red ball and is deviated to another
propagation mode. Half wave plates $H1$ and $H2$ are adjusted to rotate polarizations $V$ to $H$
and vice-versa. This causes the two beams to recombine in the second beam displacer, so that
the state at the output is the same as the state at the input.

\begin{figure*}[t]
\includegraphics[width=0.75\textwidth]{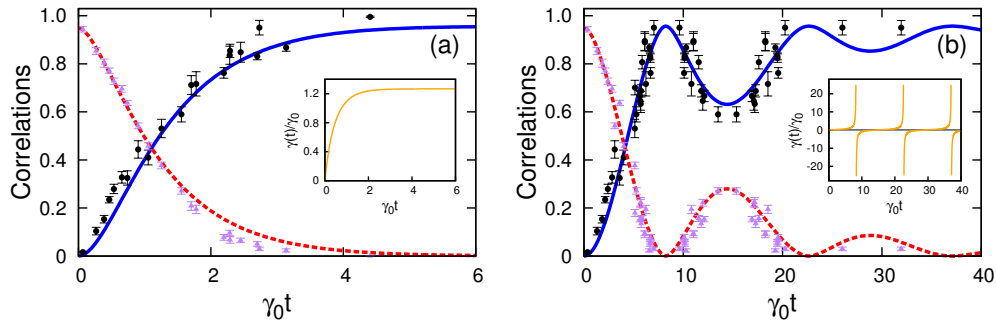}
\caption{Theoretical plot of the accessible information $J_{\mathcal{S\tilde{E}}}^\leftarrow$ (blue line) and the entanglement of formation $E_{\mathcal{S\tilde{A}}}$ (dashed red line). The insets display the decay rate $\gamma(t)/\gamma_0$ (orange line) as a function of scaled time $\gamma_0t$, that is being experimentally controlled by $p(t) \rightarrow \theta_p$. Experimental points for $J_{\mathcal{S\tilde{E}}}^\leftarrow$ and $E_{\mathcal{S\tilde{A}}}$ are shown by black dots and purple triangles, respectively. As (a) demonstrates the monotonicity of information flow in the Markovian regime with $\lambda/\gamma_0=3$, (b) displays its non-monotonous behavior in the non-Markovian regime with $\lambda/\gamma_0=0.1.$}
\label{fig6}
\end{figure*}

In Fig. \ref{fig4}c we show the case where $p \neq 0$. In this case, the half wave plate $H2$
does not rotate the polarization from $V$ to $H$ completely. There is a residual vertical component
so that the recombination of the beams in the output of the interferometer does not recover the
same state as the input. The output state has the horizontal component reduced, and this component
leaks out to the upper propagation mode. Therefore, the input mode is coupled to two output modes.
The state of this pair of modes, or path degree of freedom, represents the environment $\mathcal{E}$.
This evolution is isomorphic to an increase in the ground state component of the atom in our analogy,
or an increase in the probability of the atom to emit a photon to the environmental modes.

After propagation through the first interferometer corresponding to the application of the map,
there is a second interferometer, which is composed by beam displacers $BD2$ and $BD3$, as shown in
Fig. \ref{fig4}a. Together with the half and quarter wave plates $H4$ and $Q4$, it is used to reconstruct the state of the three qubits. The total system is given by the idler polarization, the apparatus represented by the signal photon polarization, and the environment represented by the path degree of freedom of the signal photon. Finally, for the quantum state tomography, 64 combinations of wave plate angles are used, and signal and idler photons are detected in coincidence.

\begin{figure*}[t]
\includegraphics[width=0.85\textwidth]{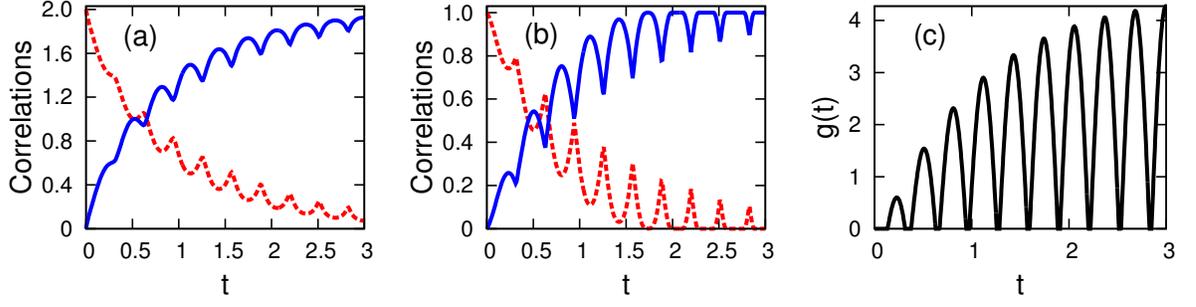}
\caption{(a) Plot of the quantum loss $\tilde{L}$ (solid blue line) and the quantum mutual information $\tilde{I}$ (dashed red line). (b) Plot of the accessible information $J_{\mathcal{S\tilde{E}}}^\leftarrow$ (solid blue line) and the entanglement of formation $E_{\mathcal{S\tilde{A}}}$ (dashed red line). (c) Plot of the the non-divisibility criterion $g(t)$. The dynamical map becomes non-divisible when $g(t)>0$. In all three plots, we set $\omega=5$ and the initial state is a maximally entangled one.}
\label{fig7}
\end{figure*}

We performed the experiment and reconstructed the final three-qubit state for different values of $p$.
From this state, we computed the correlations displayed in Fig. \ref{fig5}.
We can see the theoretical plot of the quantum loss $\tilde{L}$ (blue line) and the quantum mutual information $\tilde{I}$ (dashed red line). The curves were obtained by evolving the experimentally reconstructed initial state, which is not perfectly pure, and calculating the quantities from the evolved states. The inset displays the decay rate $\gamma(t)/\gamma_0$ (orange line) as a function of scaled time $\gamma_0t$, that is being experimentally controlled by $p(t) \rightarrow \theta_p(H2)$ the rotation angle of the half wave plate $H2$. The experimental points for $\tilde{L}$ and $\tilde{I}$ are the black dots and purple triangles, respectively. In Fig. \ref{fig5}a we observe the monotonicity of information flow in the Markovian regime with $\lambda/\gamma_0=3$. In Fig. \ref{fig5}b the non-monotonous behavior in the non-Markovian regime with $\lambda/\gamma_0=0.1$ is shown. There is a good agreement between theory and experiment.

In Fig. \ref{fig6}, we show essentially the same results as in Ref. \cite{prl}, but instead of considering an ideal pure initial state, we obtain the theoretical points from the evolution of the experimentally reconstructed initial state. We notice that taking into account this aspect leads to an improved agreement between theory and experiment. From these results we conclude that the interpretation in terms of exchange of information between $\mathcal{S}$ and $\mathcal{E}$ is valid. In the case of the amplitude damping, from Figs.\ref{fig5} and \ref{fig6} we observe that accessible information and quantum loss are essentially equivalent. In the next section, we will theoretically demonstrate that this is no longer true for the generalized amplitude damping channel.

\subsection{Generalized Amplitude Damping}

In the second example we consider the generalized amplitude channel which describes the relaxation of a quantum system when the surrounding environment is at finite temperature initially, i.e., when the environment starts from a mixed state. This phenomenological model is particularly interesting for us since it has been recently proved that the BLP measure, based on the trace distance, is not able to capture the information about the dynamics of the system coming from the non-unital parts of quantum maps \cite{nonunital}. Thus, in order to compare different approaches to information exchange, our aim here is to check the behavior of the quantum loss $\tilde{L}$ and the accessible information $J_{\mathcal{S\tilde{E}}}^\leftarrow$ under this channel, where non-divisibility actually is originated from the non-unital part of the dynamical map. In particular, the set of Kraus operators describing the dynamics of the apparatus $\mathcal{A}$ is given by
\begin{align}
 K_1(t) &= \sqrt{s(t)} \begin{pmatrix} 1 & 0\\ 0 & \sqrt{r(t)} \end{pmatrix}, \nonumber \\
 K_2(t) &= \sqrt{s(t)} \begin{pmatrix} 0 & \sqrt{1-r(t)}\\ 0 & 0 \end{pmatrix}, \nonumber \\
 K_3(t) &= \sqrt{1-s(t)} \begin{pmatrix} \sqrt{r(t)} & 0\\ 0 & 1 \end{pmatrix}, \nonumber \\
 K_4(t) &= \sqrt{1-s(t)} \begin{pmatrix} 0 & 0\\ \sqrt{1-r(t)} & 0 \end{pmatrix}
\end{align}
where $\sum_{i=1}^{4} K_i^\dagger(t) K_i(t) = I$ at all times $t$ with $s\in[0,1]$ and $r\in[0,1]$, and the quantum map $\Lambda$ represented by the above set of operators is unital, that is $\Lambda(I)=I$, if and only if $s=1/2$ or $r=1$. Similarly to what was done in \cite{nonunital}, to construct a quantum process, we choose the parameters as $s(t)=\cos^2 \omega t$ and $r(t)=e^{-t}$, where $\omega$ is a real number.

Before comparing the different approaches to monitor the information exchange between $\mathcal{S}$ and $\mathcal{E}$, let us first introduce another quantity which has been introduced in \cite{rhp}, based on the Choi-Jamiolkowski isomorphism,
\begin{equation} \label{gdiv}
g(t)=\lim_{\epsilon\rightarrow0}\frac{\rm{Tr}|(I\otimes\Lambda_{t+\epsilon,t}
)|\Omega\rangle\langle\Omega||-1}{\epsilon},
\end{equation}
where $|\Omega\rangle\langle\Omega|=\frac{1}{\sqrt{d}}\sum^{d^2-1}_{j=0}|j\rangle\otimes|j\rangle$ is the maximally entangled state of the system $\mathcal{S}$ and the apparatus $\mathcal{A}$ in the considered dimension. In fact, the condition $g(t)>0$ is a necessary and sufficient criterion for non-divisibility of dynamical quantum maps. Moreover, one can also define a measure of non-divisibility using $g(t)$ by summing it over time during the time evolution of the open system. Therefore, with the help of Eq. (\ref{gdiv}), we can investigate relation of information exchange, quantified through the quantum loss $\tilde{L}$ and the accessible information $J_{\mathcal{S\tilde{E}}}^\leftarrow$, to the regions of non-divisibility where the intermediate maps $\Lambda_{t+\epsilon,t}$ are not CPTP. For the generalized amplitude damping channel, it turns out that \cite{nonunital}
\begin{equation}
g(t)=\frac{1}{2}[|1-f(t)|+|f(t)|-1],
\end{equation}
where $f(t)=-\omega \sin(2\omega t)(1-e^{-t})+\cos^2(\omega t)$.

While we show the graphs of the quantum loss $\tilde{L}$ (solid blue line) and the quantum mutual information $\tilde{I}$ (dashed red line) in Fig. \ref{fig7}a, we display the accessible information $J_{\mathcal{S\tilde{E}}}^\leftarrow$ (solid blue line) and the entanglement of formation $E_{\mathcal{S\tilde{A}}}$ (dashed red line) in Fig. \ref{fig7}b. In all the plots, the initial state is taken as a maximally entangled one, and we set $\omega=5$. The regions of non-divisibility are displayed by the intervals where $g(t)$ is positive in Fig. \ref{fig7}c. We note that, as expected due to Eq. (\ref{dLI}) and Eq. (\ref{ej}), $\tilde{L}$ and $\tilde{I}$, and $J_{\mathcal{S\tilde{E}}}^\leftarrow$ and $E_{\mathcal{S\tilde{A}}}$ behave in an exact opposite manner. It is remarkable that, unlike the trace distance, both approaches based on information flow through entropic quantities reveal an exchange of information between the system $\mathcal{S}$ and the environment $\mathcal{E}$. However, comparing the regions with $g(t)>0$ to the intervals where $\tilde{L}$ and $J_{\mathcal{S\tilde{E}}}^\leftarrow$ temporarily decrease, we see that they do always not coincide, which is in contrast to the zero temperature relaxation model in the previous section. This model demonstrates an explicit example of how the occurrence of non-divisibility throughout the dynamics of the open system might not always imply flow of information from the environment $\mathcal{E}$ back to the system $\mathcal{S}$, even when the information exchange is measured via the quantum loss $\tilde{L}$ and the accessible information $J_{\mathcal{S\tilde{E}}}^\leftarrow$.

Furthermore, another interesting observation is that although the quantum loss $\tilde{L}$ monotonically increases until $t\approx0.5$ in Fig. \ref{fig7}a , the accessible information $J_{\mathcal{S\tilde{E}}}^\leftarrow$  decreases temporarily starting from $t\approx0.3$ in Fig. \ref{fig7}b. This clearly demonstrates that, despite their conceptual similarities, $\tilde{L}$ and $J_{\mathcal{S\tilde{E}}}^\leftarrow$ do not have to agree on the backflow of information from the environment $\mathcal{E}$ to the system $\mathcal{S}$, and can grow or decay independent of each other. Nonetheless, we note for the considered model that the accessible information $J_{\mathcal{S\tilde{E}}}^\leftarrow$ diminishes for some time in all intervals where $g(t)$ becomes positive.

\section{Conclusion}

We have presented a detailed investigation of the relation between the non-Markovianity in quantum mechanics and the flow of information between the system $\mathcal{S}$ and the environment $\mathcal{E}$. Our treatment is based on the approach of assisted knowledge where we consider a principal system $\mathcal{S}$ that is initially correlated with its measurement apparatus $\mathcal{A}$. Although there is no direct interaction between the system $\mathcal{S}$ and the environment $\mathcal{E}$, there still exists an exchange of information among the constituents of the tripartite system $\mathcal{SAE}$, due to the fact that the apparatus $\mathcal{A}$ interacts with the environment $\mathcal{E}$. Centered on this scenario, we have introduced a new way of understanding the information exchange between the system $\mathcal{S}$ and the environment $\mathcal{E}$ through the quantum loss $\tilde{L}$, which quantifies the amount of residual information that the environment $\mathcal{E}$ and the system $\mathcal{S}$ have in common after the interaction. We have also shown how measuring the information flow and thus non-Markovianity via quantum loss is in fact equivalent to utilizing the LFS measure of non-Markovianity. This equivalence gives a straightforward information theoretic interpretation to the LFS measure. Moreover, recognizing that using the entanglement-based RHP measure is equivalent to the accessible information approach, we have provided an alternative way of quantifying the exchange of information between the system $\mathcal{S}$ and the environment $\mathcal{E}$. More important, we have also revealed a clear connection between two apparently unrelated measures of non-Markovianity, namely the LFS and the RHP measures, by making use of the link between the quantum loss $\tilde{L}$ and the accessible information $J_{\mathcal{S\tilde{E}}}^\leftarrow$. In particular, the only conceptual difference between these two quantities lies on the local accessibility of the information shared between the system $\mathcal{S}$ and the environment $\mathcal{E}$, when local observations are performed on the environment $\mathcal{E}$.

We have studied the information exchange in terms of the quantum loss $\tilde{L}$ and the accessible information $J_{\mathcal{S\tilde{E}}}^\leftarrow$ in two paradigmatic models, namely for the zero and finite temperature relaxation processes. For the zero temperature case, we have demonstrated that both the quantum loss $\tilde{L}$  and the accessible information $J_{\mathcal{S\tilde{E}}}^\leftarrow$ are able to capture the flow of information in a similar way. Moreover, we have provided an experimental simulation of this process using an all optical setup that allows full access to the environment. Our experimental results are shown to be in good agreement with the theoretical predictions. For the finite temperature relaxation model, we have explored the similarities and the differences of measuring information flow in terms of the trace distance, the quantum loss $\tilde{L}$  and the accessible information $J_{\mathcal{S\tilde{E}}}^\leftarrow$. Specifically, we have shown that, while the trace distance fails to capture the inverse flow of information originated from the non-unital part of the dynamical quantum map, both the quantum loss $\tilde{L}$  and the accessible information $J_{\mathcal{S\tilde{E}}}^\leftarrow$ can successfully identify the exchange of information between the system $\mathcal{S}$ and the environment $\mathcal{E}$ in this case. On the other hand, we have also found that, despite their conceptual similarities, it is possible for the quantum loss (the LFS measure) and the accessible information (the RHP measure) to disagree on the flow of information in certain time intervals during the time evolution.

\begin{acknowledgments}
Financial support was provided by Brazilian agencies CNPq, CAPES, FAPERJ, FAPESP, and the National Institute for Quantum Information, INCT-IQ.
\end{acknowledgments}

\end{document}